\documentclass[prb,twocolumn,reprint,amsmath,amssymb,superscriptaddress,floatfix]{revtex4}
\usepackage[breaklinks=true,colorlinks,citecolor=blue,linkcolor=blue,urlcolor=blue]{hyperref}
\usepackage{float}
\usepackage{epsfig,mathrsfs,color,latexsym,subfigure,marginnote}

\usepackage{graphicx}  
\usepackage{dcolumn}   
\usepackage{bm}        
\usepackage{amssymb}   
\usepackage{siunitx}
 
\usepackage{color} 
\newcommand{\I}{\mathcal{I}}
\newcommand{\T}{\mathcal{T}}

\begin{document}

\title{Topological crystalline insulator state with type-II Dirac fermions in transition metal dipnictides}

\author{Baokai Wang}
\affiliation{Department of Physics, Northeastern University, Boston, Massachusetts 02115, USA}

\author{Bahadur Singh$^*$}
\affiliation{Department of Physics, Northeastern University, Boston, Massachusetts 02115, USA}
\affiliation{SZU-NUS Collaborative Center and International Collaborative Laboratory of 2D Materials for Optoelectronic Science $\&$ Technology, Engineering Technology Research Center for 2D Materials Information Functional Devices and Systems of Guangdong Province, Institute of Microscale Optoelectronics, Shenzhen University, Shenzhen, 518060, China}

\author{Barun Ghosh}
\affiliation{Department of Physics, Indian Institute of Technology Kanpur, Kanpur 208016, India}

\author{Wei-Chi Chiu}
\affiliation{Department of Physics, Northeastern University, Boston, Massachusetts 02115, USA}

\author{M. Mofazzel Hosen}
\affiliation{Department of Physics, University of Central Florida, Orlando, Florida 32816, USA}

\author {Qitao Zhang}
\affiliation{SZU-NUS Collaborative Center and International Collaborative Laboratory of 2D Materials for Optoelectronic Science $\&$ Technology, Engineering Technology Research Center for 2D Materials Information Functional Devices and Systems of Guangdong Province, Institute of Microscale Optoelectronics, Shenzhen University, Shenzhen, 518060, China}

\author{Li Ying}
\affiliation{SZU-NUS Collaborative Center and International Collaborative Laboratory of 2D Materials for Optoelectronic Science $\&$ Technology, Engineering Technology Research Center for 2D Materials Information Functional Devices and Systems of Guangdong Province, Institute of Microscale Optoelectronics, Shenzhen University, Shenzhen, 518060, China}

\author{Madhab Neupane}
\affiliation{Department of Physics, University of Central Florida, Orlando, Florida 32816, USA}

\author{Amit Agarwal}
\affiliation{Department of Physics, Indian Institute of Technology Kanpur, Kanpur 208016, India}

\author{Hsin Lin}
\affiliation{Institute of Physics, Academia Sinica, Taipei 11529, Taiwan}

\author{Arun Bansil\footnote{Corresponding authors' emails: bahadursingh24@gmail.com,  ar.bansil@northeastern.edu}}
\affiliation{Department of Physics, Northeastern University, Boston, Massachusetts 02115, USA}


\begin{abstract}
The interplay between topology and crystalline symmetries in materials can lead to a variety of topological crystalline insulator (TCI) states. Despite significant effort towards their experimental realization, so far only Pb$_{1-x}$Sn$_x$Te has been confirmed as a mirror-symmetry protected TCI. Here, based on first-principles calculations combined with a symmetry analysis, we identify a rotational-symmetry protected TCI state in the transition-metal dipnictide RX$_2$ family, where R = Ta or Nb and X = P, As, or Sb. Taking TaAs$_2$ as an exemplar system, we show that its low-energy band structure consists of two types of bulk nodal lines in the absence of spin-orbit coupling (SOC) effects. Turning on the SOC opens a continuous bandgap in the energy spectrum and drives the system into a $C_2T$-symmetry-protected TCI state. On the (010) surface, we show the presence of rotational-symmetry-protected nontrivial Dirac cone states within a local bulk energy gap of $\sim$ 300 meV. Interestingly, the Dirac cones have tilted energy dispersion, realizing a type-II Dirac fermion state in a topological crystalline insulator. Our results thus indicate that the TaAs$_2$ materials family provides an ideal setting for exploring the unique physics associated with type-II Dirac fermions in rotational-symmetry-protected TCIs.
\end{abstract}

\maketitle
\section {Introduction}
Topological insulators (TIs) represent a new state of quantum matter which is described by the topology of the bulk bands instead of a local order parameter within the Landau paradigm \cite{Bansil:2016, Hasan:2010, Zhang:2011}. These materials support an odd number of metallic surface states with linear energy dispersion while remaining insulating in the bulk. The topological surface states are protected by time-reversal symmetry and are immune to nonmagnetic impurities. Soon after the realization of TIs, the topological classification of insulating electronic structures was extended beyond the time-reversal symmetry protected states to encompass crystalline symmetries. The topological nature of topological crystalline insulators (TCIs) arises from the bulk crystal symmetries \cite{Liang:2011}. Owing to the richness of crystal symmetries many possible TCI states are possible. The first TCI state was predicted theoretically in the SnTe materials class \cite{Hsieh:2012}, which was subsequently verified in experiments by observing surface Dirac-cone states\cite{Dziawa:2012topological, Xu:2012observation, Tanaka2012:experimental}. The nontrivial topology in SnTe appears due to the presence of mirror-symmetry and it is manifested by the existence of an even number of Dirac-cone states over the surface.

Recently a new TCI state protected by $N$-fold rotational symmetries was proposed in time-reversal-invariant systems with spin-orbit coupling (SOC)\cite{Fang:2017rtci,Song:2017rtci,HOTI_Titus,Cheng:2018}. Such rotational-symmetry-protected TCIs are distinct from mirror-symmetry protected TCIs and support $N$ Dirac cones on the surface normal to the rotational axis. Their Dirac cone states are not restricted to high-symmetry points and can appear at generic $k$ points in the Brillouin zone (BZ).
Also, the rotational-symmetry-protected TCIs evade the fermion multiplication theorem to drive a rotational anomaly, and harbor helical edge states on the hinges of their surfaces parallel to the rotational axis. This class of TCIs can support anomalous transport properties and it could provide a basis for realizing Majorana zero modes through proximity induced superconductivity\cite{Fang:2017rtci,Song:2017rtci,HOTI_Titus,Cheng:2018,Majorana_Andrei}.

Monoclinic lattices with $C_{2h}$ point-group symmetry are ideally suited in connection with materials discovery of rotational-symmetry protected TCIs \cite{Fang:2017rtci,Song:2017rtci,HOTI_Titus,Song:2018,Hsu_2019}. $C_{2h}$ symmetry includes a mirror plane $M_{[0 1 0]}$, a two-fold rotational axis $C_{2[0 1 0]}$, and the space inversion symmetry $\I$. Although this symmetry group can support both a mirror-symmetry protected TCI (Fig. \ref{fig:CS}(a)) as well as a rotational-symmetry protected TCI (Fig. \ref{fig:CS}(b)), the associated gapless surface states are located on different surfaces. In particular, the (010) surface preserves the two-fold rotational symmetry $C_{2[0 1 0]}$, and therefore this surface can support rotational-symmetry protected surface states. Rotational-symmetry-protected TCIs have been predicted recently in ${\alpha-\rm Bi_4Br_4}$ \cite{Hsu_2019}, $\rm Ca_2As$ \cite{Zhou:2018rtci}, and Bi \cite{Hsu2019_Bi}.  

In this paper, we discuss the existence of a rotational-symmetry-protected TCI phase in the transition-metal dipnictides RX$_2$ (R = Ta or Nb and X = P, As, or Sb) materials class with $C_{2h}$ lattice symmetries. RX$_2$ materials in which many intriguing properties are observed have been realized in experiments. For example, ${\rm NbAs_2}$ shows a large, non-saturating transverse magnetoresistance and a negative longitudinal magnetoresistance, which may be reflective of its non-trivial bulk band topology  \cite{Shen:2016}. Other experiments show that at low temperatures, thermal conductivity of  TaAs$_2$ scales with temperature as $T^4$, while the resistivity is independent of $T$, indicating breakdown of the Weidemann-Franz law and possible presence of a non-Fermi liquid state \cite{rao2019quantum}. Notably, the RX$_2$ materials have been predicted as nearly electron-hole compensated semimetals in which a continuous SOC driven bandgap between the valence and conduction bands leads to weak topological invariants $(\nu_0; \nu_1\nu_2\nu_3) = (0; 111)$\cite{Xu:2016weak, Luo:2016anomalous, Gresch:2017}. Regardless, to the best of our knowledge, a rotational-symmetry-protected TCI phase has not been discussed previously in the literature in the RX$_2$ materials family. 

Our analysis reveals that the RX$_2$ materials realize the $C_{2[010]}$ rotational-symmetry-protected TCI state. Taking TaAs$_2$ as an example, we show in-depth that it supports two types of nodal lines in the absence of the SOC effects. Inclusion of the SOC gaps out the nodal lines and drives the system into a topological state with weak topological invariants (0;111) and symmetry indicators $(Z_2Z_2Z_2; Z_4)=(111;2)$. A careful inspection of the topological state shows that $\rm TaAs_2$ harbors a $C_2T$ symmetry-protected TCI state. To highlight the associated nontrivial band topology, we present the (010)-surface electronic spectrum and show the existence of rotational-symmetry-protected nontrivial Dirac cone states at generic $k$ points within a local bulk energy gap of $\sim$300 meV. The Dirac cones are found to exhibit a unique type-II energy dispersion. In this way, our study demonstrates that transition metal dipnictides RX$_2$ could provide an experimentally viable platform for exploring rotational-symmetry-protected TCIs with type-II Dirac cones.

The organization of the remainder of the paper is as follows. In Sec. \ref{method}, we provide the methodology and structural details of the $\rm{RX_2}$ compounds. The bulk topological electronic structure is explored in Sec. \ref{bulk_bands}. In Sec. \ref{surface_bands}, we discuss surface electronic structure and $C_2T$ symmetry protected Dirac cone states in TaAs$_2$. Finally, we summarize conclusions of our study in Sec. \ref{disc}.

\section{Computational details}\label{method}

Electronic structure calculations were performed within the framework of the density functional theory (DFT) using the projector augmented wave (PAW) method as implemented in the VASP suite of codes\cite{kohan_dft,Kresse:1993, Kresse:199615, Kresse:1996prb, Kresse:1999paw}. The generalized-gradient-approximation (GGA) was used to incorporate exchange-correlation effects\cite{PBE}. An energy cutoff of 350 eV was used for the plane-wave basis set and a $\Gamma$-centered 12$\times$12$\times$8 $k$-mesh was used for BZ integrations. We started with experimental lattice parameters and relaxed atomic positions until the residual forces on each atom were less than 0.001 eV/\AA.   We constructed a tight-binding model with atom-centered Wannier functions using the VASP2WANNIER90 interface\cite{W90}. The surface energy spectrum was obtained within the iterative Green's function method using the Wanniertools package\cite{Wu:2017}.

\begin{figure}
\includegraphics[width = 0.5\textwidth]{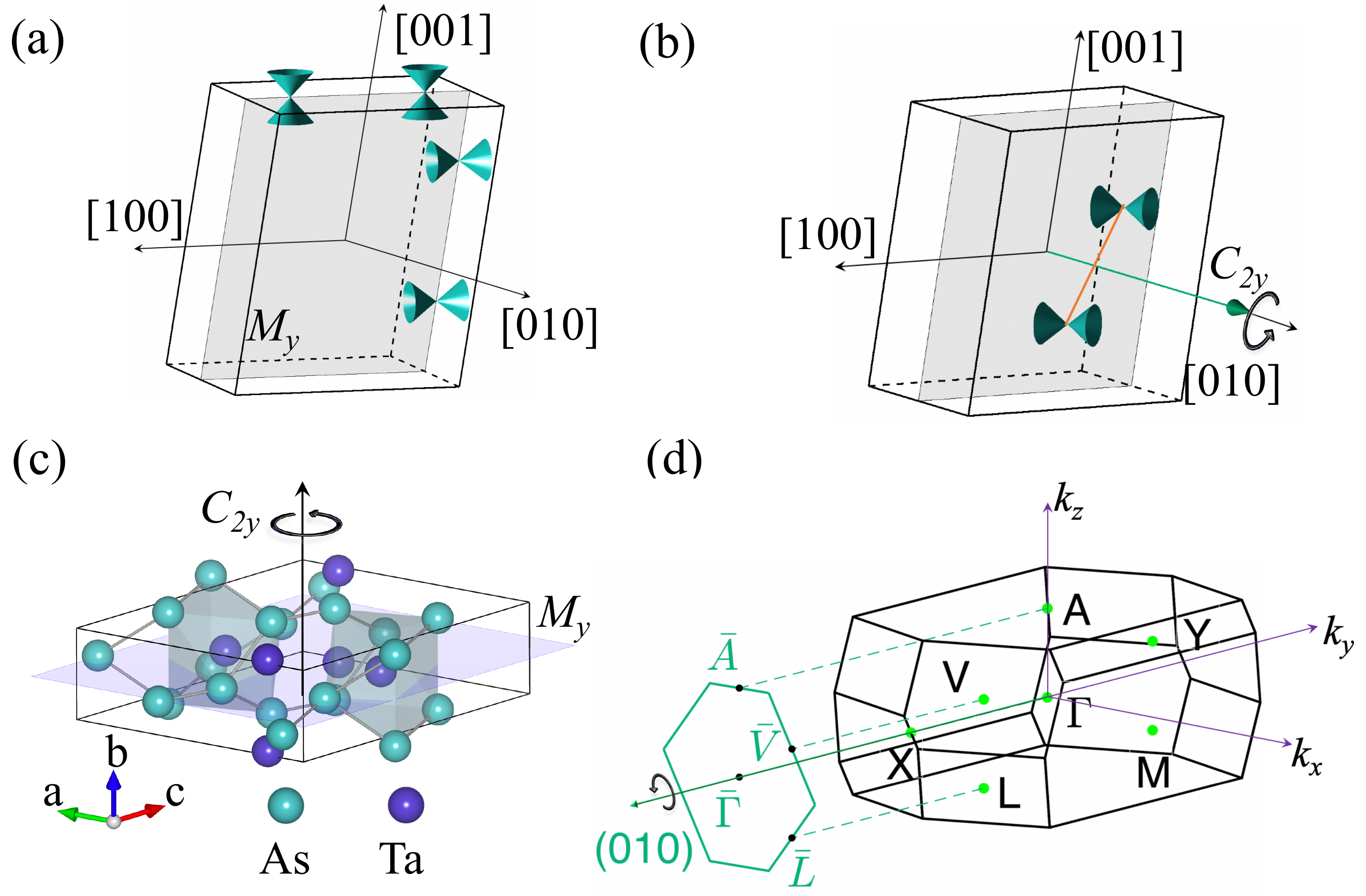}
\caption{Schematic illustration of the Dirac-cone surface states of (a) a mirror-symmetry protected topological crystalline insulator (TCI)  and (b) a $C_{2[010]}$ rotational-symmetry protected TCI with $C_{2h}$ point-group symmetry. The (010) mirror plane ($M_y$, shaded gray plane) and high-symmetry crystal axes are shown. The Dirac cones in a mirror-symmetry protected TCI are pinned to the mirror-invariant line over the surface whereas they lie at generic $k$ points on the surface normal to the rotational axis in a rotational-symmetry-protected TCI. (c) The conventional unit cell of RX$_2$ compounds. The $M_y$ (010) mirror-invariant plane and $C_{2y}$ ($C_{2[010]}$) rotational axis are shown. (d) The primitive Brillouin zone and its projection on the (010) surface.}
\label{fig:CS}
\end{figure}

Transition metal dipnictides RX$_2$ crystallize in a monoclinic Bravais lattice with space-group $C2/m$ (No. 12)  \cite{Wang:2014nbsb2, Li:2016tasb2, Wang:2016nbas2,Wu:2016apl, Rundqvist:1966new}. The crystal structure is shown in Fig. \ref{fig:CS}(c). The primitive unit consists of two transition-metal (R) atoms and four pnictogen (X) atoms.   This crystal structure supports a two-fold rotational axis $C_{2[010]}$, mirror-plane symmetry $M_{[010]}$, and inversion symmetry $\I$. The RX$_2$ materials are nonmagnetic and respect time-reversal symmetry $\T$. The bulk BZ and the associated (010) surface BZ are shown in Fig. 1(d) where the high-symmetry points are indicated.

\begin{figure}
\includegraphics[width = 0.5\textwidth]{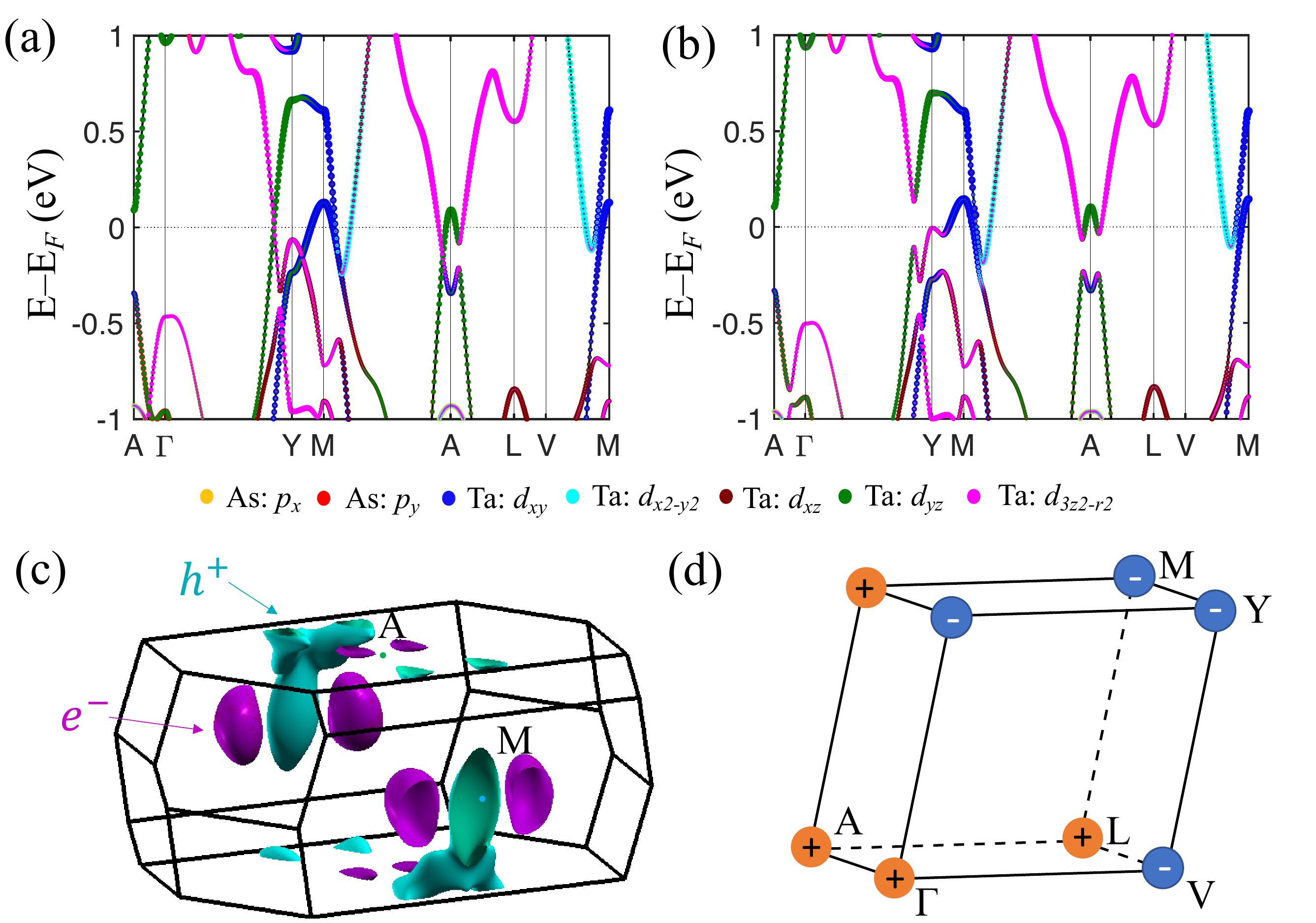}
\caption{Bulk band structure of TaAs$_2$ (a) without and (b) with spin-orbit coupling (SOC). The orbital compositions of bands are shown using various colors. The band crossings in (a) are seen resolved along the $\Gamma-Y$ and $M-A$ directions; these are gapped in the presence of the SOC. (c) Fermi surface of TaAs$_2$ with electron (cyan) and hole (purple) pockets. (d) Parity eigenvalues of the valence bands at eight time-reversal-invariant momentum points in the BZ.}
\label{fig:BS}
\end{figure}

\section{Bulk electronic structure and topological invariants}\label{bulk_bands}

The bulk band structure of TaAs$_2$ without and with including SOC is presented in Figs. \ref{fig:BS}(a) and \ref{fig:BS}(b), respectively. The orbital character of the bands (color coded) shows that bands near the Fermi level mainly arise from the Ta $d$ and As $p$ states. Without the SOC, the valence and conduction bands are seen to cross along the $\Gamma-Y$ and $M-A$ high-symmetry directions. On the inclusion of SOC, a bandgap opens up at the band-crossing points, separating the valence and conduction bands locally at each $k$ point. This separation leads to well-defined band manifolds and facilitates the calculation of topological invariants as in the insulators. Since the $\rm TaAs_2$ crystal respects inversion symmetry, it is possible to calculate the $Z_2$ invariants $(\nu_0; \nu_1\nu_2\nu_3) $ from the parity eigenvalues of the valence bands at the time-reversal-invariant momentum points \cite{fu_parity}. In Fig. \ref{fig:BS}, we present these results and find $Z_2$ invariants as $(0; 111)$. These agree well with the earlier studies and indicate that TaAs$_2$ is a weak TI\cite{Xu:2016weak, Luo:2016anomalous, Gresch:2017}.
 
We emphasize that despite the opening of a local bandgap between the valence and conduction states, $\rm TaAs_2$ preserves its semimetal character with the presence of electron and hole pockets. This is seen clearly in the Fermi surface plot of Fig. \ref{fig:BS}(c). We find one hole and four electron pockets in the BZ. The volume of the electron and hole pockets is roughly the same, indicating that TaAs$_2$ is nearly an electron-hole compensated semimetal. This feature of the electronic spectrum could drive the large, nonsaturating transverse magnetoresistance observed experimentally in $\rm TaAs_2$ \cite{Shen:2016}.

\begin{figure}
\includegraphics[width = 0.5\textwidth]{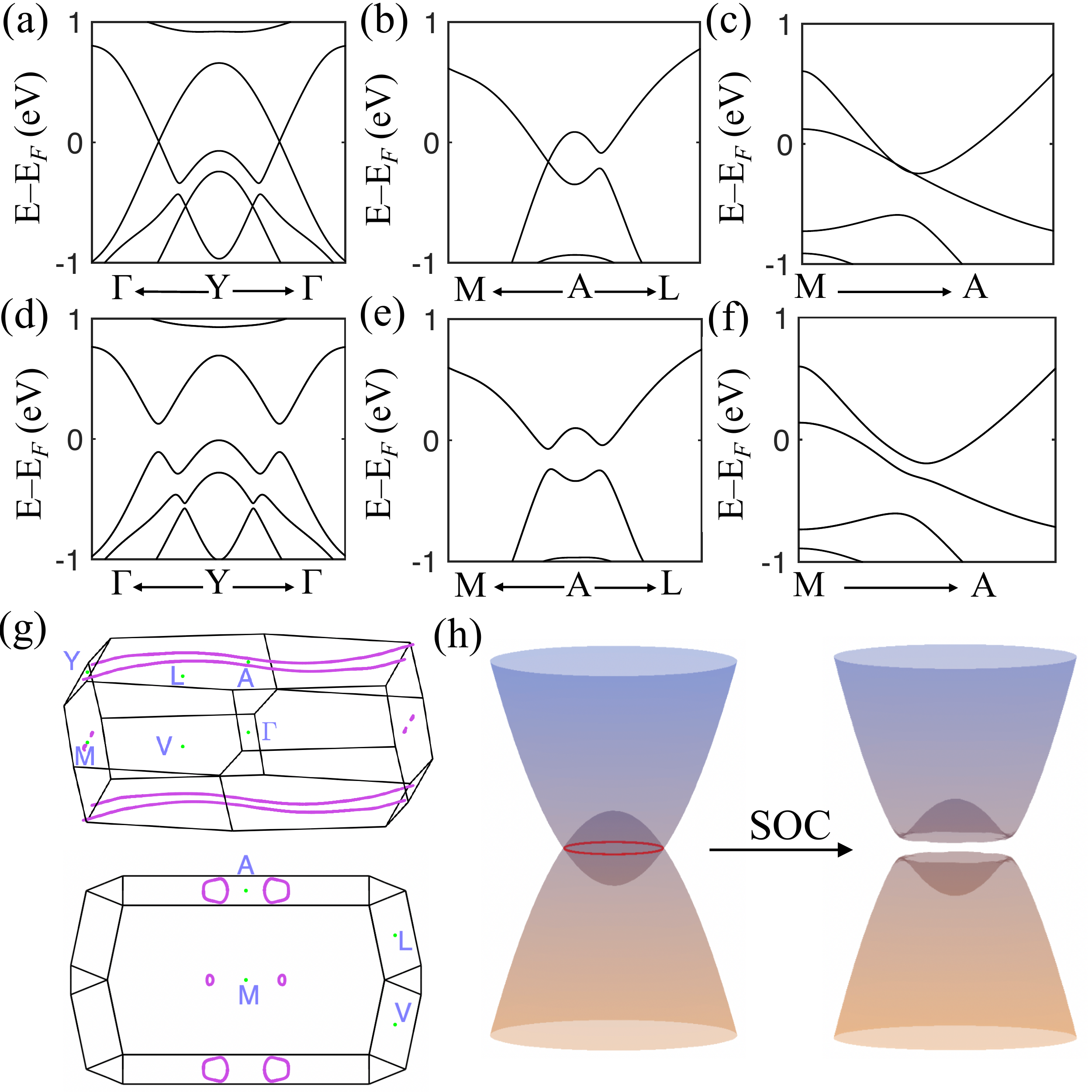}
\caption{ (a)-(f) Bulk band structure of $\rm TaAs_2$ around the selected time-reversal-invariant momentum points $Y$, $A$, and $M$ where band inversion takes place. The top row shows band structure without SOC whereas the middle row shows band structure including SOC. (g) Two different views of the nodal-line structures in the BZ. The top figure highlights nodal lines extending across the BZs and the bottom figure highlights nodal rings formed near the $M$ point. (h) Schematic illustration of the formation of nontrivial insulating states with SOC.}
\label{fig:TCI}
\end{figure}

In order to characterize the nodal-line and TCI states of TaAs$_2$, we examine the band-crossings in Fig. \ref{fig:TCI}. In the absence of SOC, there are band crossings along the $\Gamma-Y$ and $M-A$ symmetry lines, see Figs. \ref{fig:TCI}(a)-(c). A careful inspection shows that these band-crossings form Dirac nodal lines in the BZ, see Fig. \ref{fig:TCI}(g). There are two types of nodal lines. The first type includes two non-closed spiral nodal-lines extending across the BZ through point $A$ whereas the second type includes two nodal-loops near the $M$ point. When SOC is included, these nodal lines are gapped (see Figs. \ref{fig:TCI}(d)-(f)), and lead to a band inversion at the $Y$ and $A$ points. The band inversion at these points primarily involves Ta $d_{yz}$ and $d_{z^2}$ orbitals. Since the SOC separates valence and conduction states by a local bandgap, the symmetry indicators for identifying specific topological states for gapped systems now become well defined \cite{Song:2018, Tang:2019, Po:2017,ChenFang_database,Andrei_database,Viswanath_database}.  In particular, the symmetry indicators are obtained from the full set of eigenvalues of the space-group symmetry operators of the occupied bands at high-symmetry points. Following Ref.\cite{Song:2018}, the topological phase in space group No. $\#12$ is described by a set of four numbers $(Z_2Z_2Z_2; Z_4)$. The computed symmetry indicators and topological invariants for TaAs$_2$ are listed in Table 1. 
The results of Table 1 show that the topological state of TaAs$_2$ is described by weak topological invariants (111), non-zero rotational invariant $n_{2^{010}}=1$, glide invariant $n_{g^{010}_{\bar{\frac{1}{2}}00}}=1$ and inversion invariant $n_i=1$ \footnote {Note that we have only considered the rotational symmetry protected topological states on the (010) surface of TaAs$_2$ in Sec. \ref{surface_bands}. Exploration of other nonzero invariants will be interesting.}.
 
 \begin{table}[h!]
\caption{Calculated symmetry indicators and topological invariants for $\rm TaAs_2$. $(\nu_0;\nu_1,\nu_2,\nu_3)$ are the $Z_2$ invariants for a three-dimensional TI, $n_{m^{010}}$ is the mirror-chern number for (010) plane, and $n_{2^{010}}$ is an invariant for the two-fold rotational symmetry $C_{2[010]}$. $n_{g^{010}_{\bar{\frac{1}{2}}00}}$, $n_i$ and $n_{2^{010}_1}$ are topological invariants associated with glide symmetry, inversion symmetry and screw symmetry,  respectively. }

\renewcommand{\arraystretch}{1.75}
\begin{tabular}{c c c c c c c c}
\hline \hline 
 $(\mathcal{Z}_2,\mathcal{Z}_2,\mathcal{Z}_2,\mathcal{Z}_4)$ & $(\nu_0;\nu_1,\nu_2,\nu_3)$ & $n_{m^{010}}$ & $n_{2^{010}}$ & $n_{g^{010}_{\bar{\frac{1}{2}}00}}$ & $n_i$ & $n_{2^{010}_1}$ \\ 
\hline 
 (1,1,1,2) & (0;111) & 0& 1 & 1 & 1 & 0\\ 
\hline  \hline
\end{tabular} 
\label{T1}
\end{table} 

\section{Surface electronic structure}\label{surface_bands}

\begin{figure}[ht!]
\includegraphics[width = 0.5\textwidth]{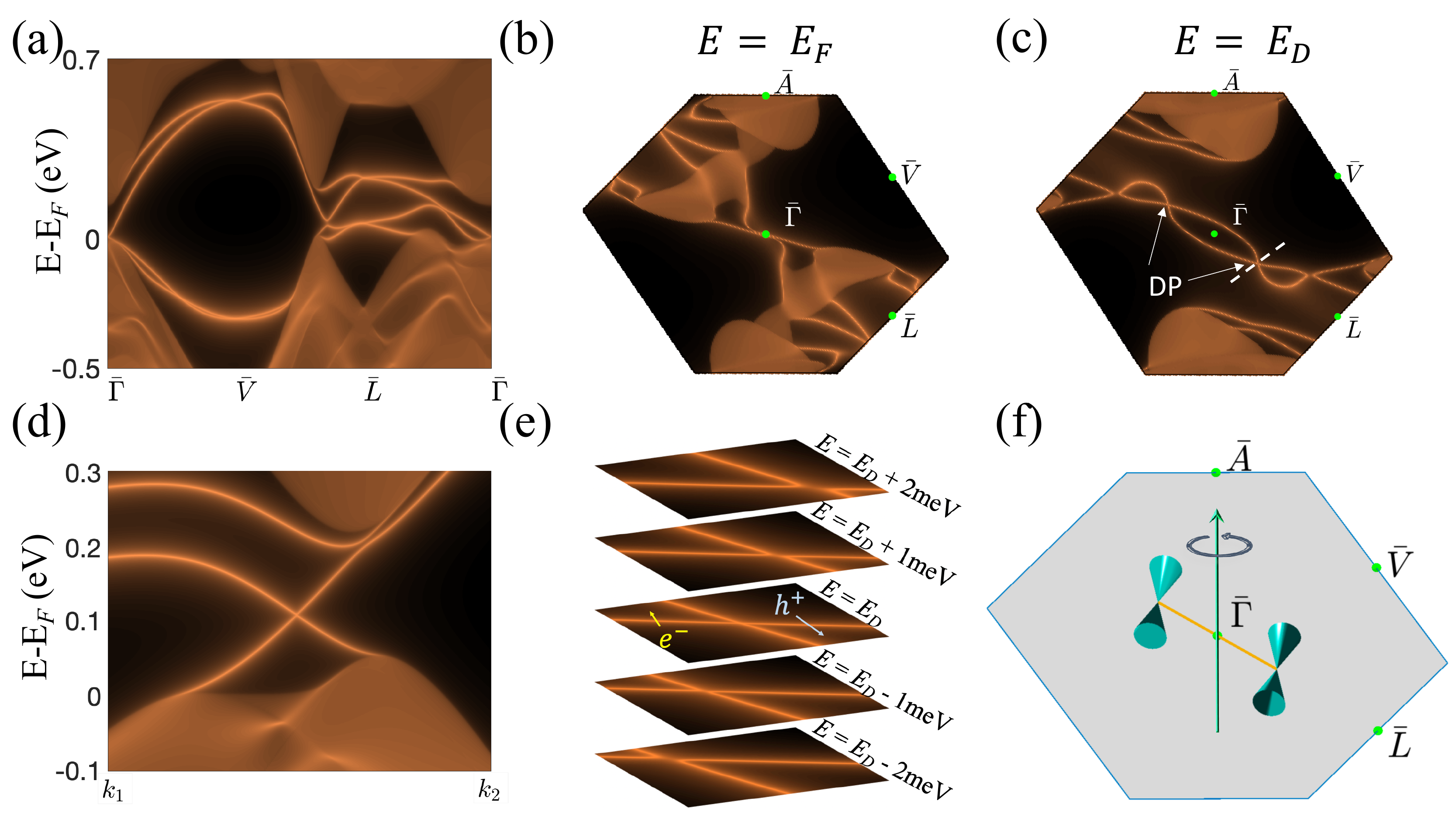}
\caption{(a) Surface band structure of $\rm TaAs_2$ along the high-symmetry lines in the (010) surface BZ. Constant energy contours at (b) $E = E_f$ and (c) $E = E_D $, where $E_D = 110$ meV denotes the energy of the Dirac point (DP). Dirac points are marked with white arrows. (d) Band structure along the $k$-path marked with a dashed white line in (c). (e) Constant energy contours near the DP. The electron and hole pockets are noted in the plots. (f) Schematic illustration of the rotational-symmetry protected Dirac cone states on the (010) surface of $\rm TaAs_2$. }
\label{fig:SS}
\end{figure}

The nontrivial value of $n_{2^{010}}$ dictates that the (010) surface normal to the two-fold $C_{2[010]}$ rotational axis supports two Dirac-cone state. In order to showcase these states, we present the (010) surface band structure in Fig. \ref{fig:SS}, where a topological surface state centered at the $\overline{\Gamma}$ point connecting the bulk valence and conduction bands can be seen clearly in Fig.  \ref{fig:SS}(a), confirming the nontrivial topological nature of the material. The associated Fermi surface contours are shown in Fig.  \ref{fig:SS}(b). The Dirac-cone states associated with rotational-symmetry-protected TCIs generally lie at generic $k$ points on the rotationally invariant surface. Therefore, we scanned the entire (010) surface BZ and found these cones to lie at (0.1248, -0.0974) $\r{AA}^{-1}$ and (-0.1248, 0.0974) $\r{AA}^{-1}$ close to the $\overline{\Gamma}-\overline{L}$ direction as seen in the constant energy contours of Fig. \ref{fig:SS}(c). Energy dispersion of these Dirac cone states along the path marked by the white dashed line in Fig. \ref{fig:SS}(c) is shown in Fig. \ref{fig:SS}(d). The states lie within a local bandgap of $\sim$ 300 meV and exhibit a tilted type-II energy dispersion as seen from the constant-energy contours near the Dirac points in Fig.  \ref{fig:SS}(e). The Dirac cones appear at the touching point between the electron and hole pockets. At energies below the Dirac cone energy $E_D$, the size of the hole pockets increases while that of the electron pockets shrinks. This behavior is reversed as we go to energies above $E_D$. These results clearly indicate that TaAs$_2$ harbors a type-II rotational-symmetry-protected Dirac cone state as schematically shown in Fig.  \ref{fig:SS}(f).

\section {Summary and conclusions}\label{disc}

The type-II Dirac cone state we have delineated in this study will be interesting for exploring exotic properties of TCIs. Notably, type-II Dirac cone states are prohibited in strong topological insulators with time-reversal symmetry\cite{Chiu_TCI}. In contrast, they have been predicted in the mirror-symmetry protected TCI family of antipervoskites \cite{Chiu_TCI}. Such type-II Dirac cones exhibit characteristic van Hove singularities in their density of surface states and have a Landau level spectrum that is distinct from type I Dirac states, leading to unique electronic and magnetotransport properties\cite{Singh18_Saddle,Ghosh_HgPt2Se3}. Since $\rm TaAs_2 $ has been realized experimentally, our theoretically predicted type-II Dirac cone states in this material and the associated nontrivial properties would be amenable to experimental verification.  

Our calculations on the entire family of transition metal dipnictides $\rm RX_2$ show that their band structure closely resembles that of $\rm TaAs_2$ (see the Appendix 
\ref{app:A} for details).  This family of materials supports a nodal-line structure without the SOC and transitions to a TCI state with SOC. The symmetry indicators and topological invariants reveal that these compounds are characterized by weak topological invariants (111) and a nonzero rotational invariant $n_{2^{010}}$= 1. The $\rm RX_2$ family thus realizes the rotational-symmetry-protected TCI state. 

In conclusion, we have identified the presence of a rotational-symmetry-protected TCI state in the transition-metal dipnictides materials family. Taking $\rm TaAs_2$ as an exemplar system, we show that it supports a nodal-line semimetal state with two distinct nodal lines without the SOC. Inclusion of the SOC gaps out the nodal-lines and transitions the system into a topological state with well-separated valence and conduction band manifolds.  The symmetry indicators and topological invariants reveal that $\rm TaAs_2$ has symmetry indicator $(Z_2Z_2Z_2; Z_4)= (111;2)$ and is a $C_{2[010]}$ rotational-symmetry-protected TCI. We further confirmed this TCI state by calculating the (010)-surface band structure, which shows the existence of type-II Dirac-cone states in a local bulk energy gap of $~300$ meV. Our study thus shows that the $\rm TaAs_2$ class of transition metal dipnictides would provide an excellent platform for exploring TCIs with type-II Dirac fermions and the associated nontrivial properties.

\section*{ACKNOWLEDGEMENTS}

The work at Northeastern University was supported by the US Department of Energy (DOE), Office of Science, Basic Energy Sciences Grant No. DE-FG02-07ER46352, and benefited from Northeastern University’s Advanced Scientific Computation Center and the National Energy Research Scientific Computing Center through DOE Grant No. DE-AC02-05CH11231. The work at Shenzhen University was supported by the Shenzhen Peacock Plan (Grant No. KQTD2016053112042971) and the Science and Technology Planning Project of Guangdong Province (Grant No. 2016B050501005). H.L. acknowledges Academia Sinica, Taiwan for support under Innovative Materials and Analysis Technology Exploration (AS-iMATE-107-11). BG acknowledges CSIR for the Senior Research Fellowship. The work at IIT Kanpur was benefited from the high-performance facilities of the computer center of IIT Kanpur. M.N. is supported by the Air Force Office of Scientific Research under award number FA9550-17-1-0415 and the National Science Foundation (NSF) CAREER award DMR-1847962.

\bibliographystyle{prsty_SM}
\bibliography{ref}

\appendix 

\section{Band structure of transition metal dipnictide materials family}
\label{app:A}

\begin{figure}[h!]
\includegraphics[width = 0.5\textwidth]{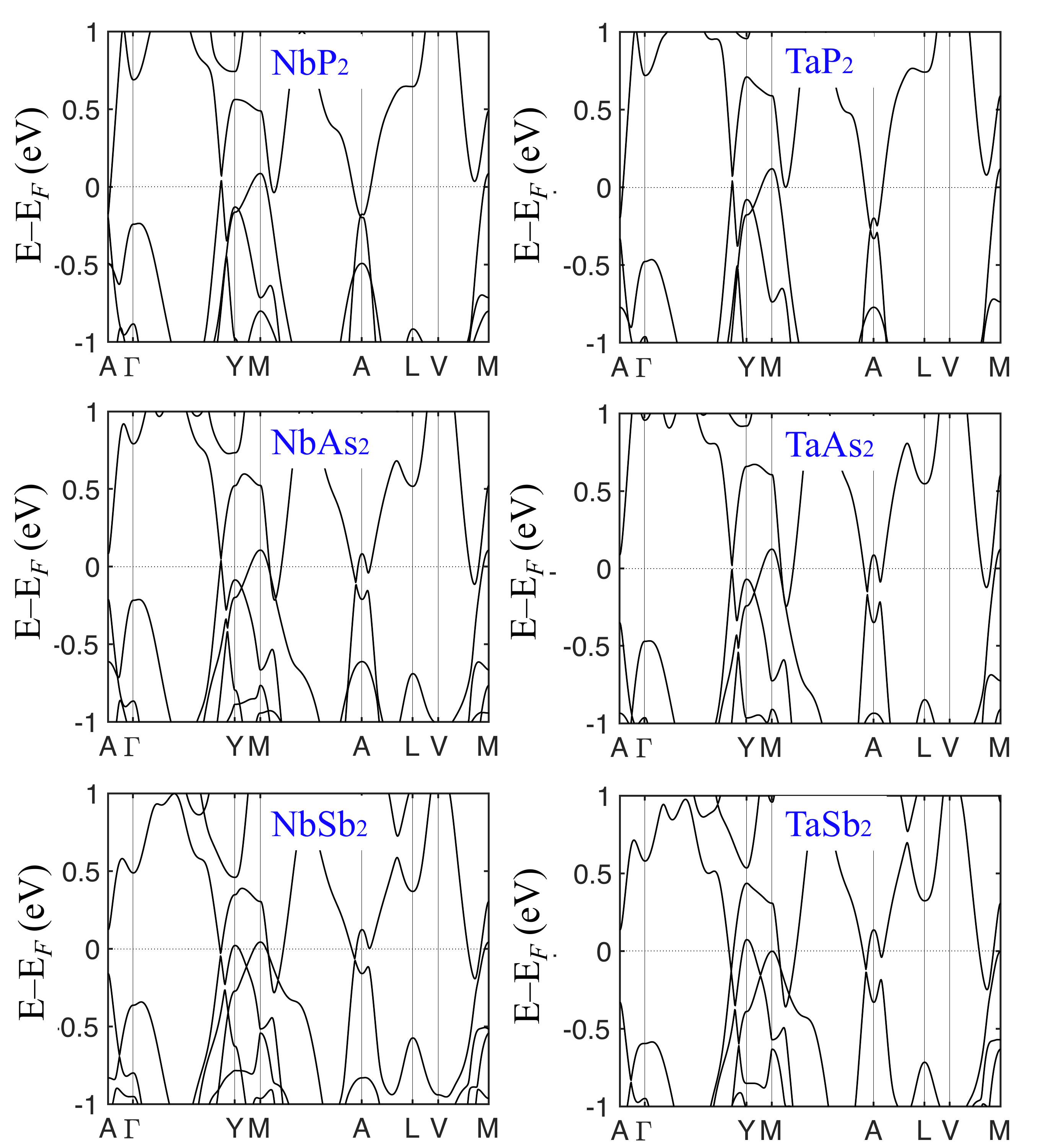}
\caption{Calculated bulk band structures of the transition metal dipnictide family of materials without the SOC. Blue text on each figure identifies the material.}
\label{bs_nosoc}
\end{figure}

\begin{figure}[ht!]
\includegraphics[width = 0.5\textwidth]{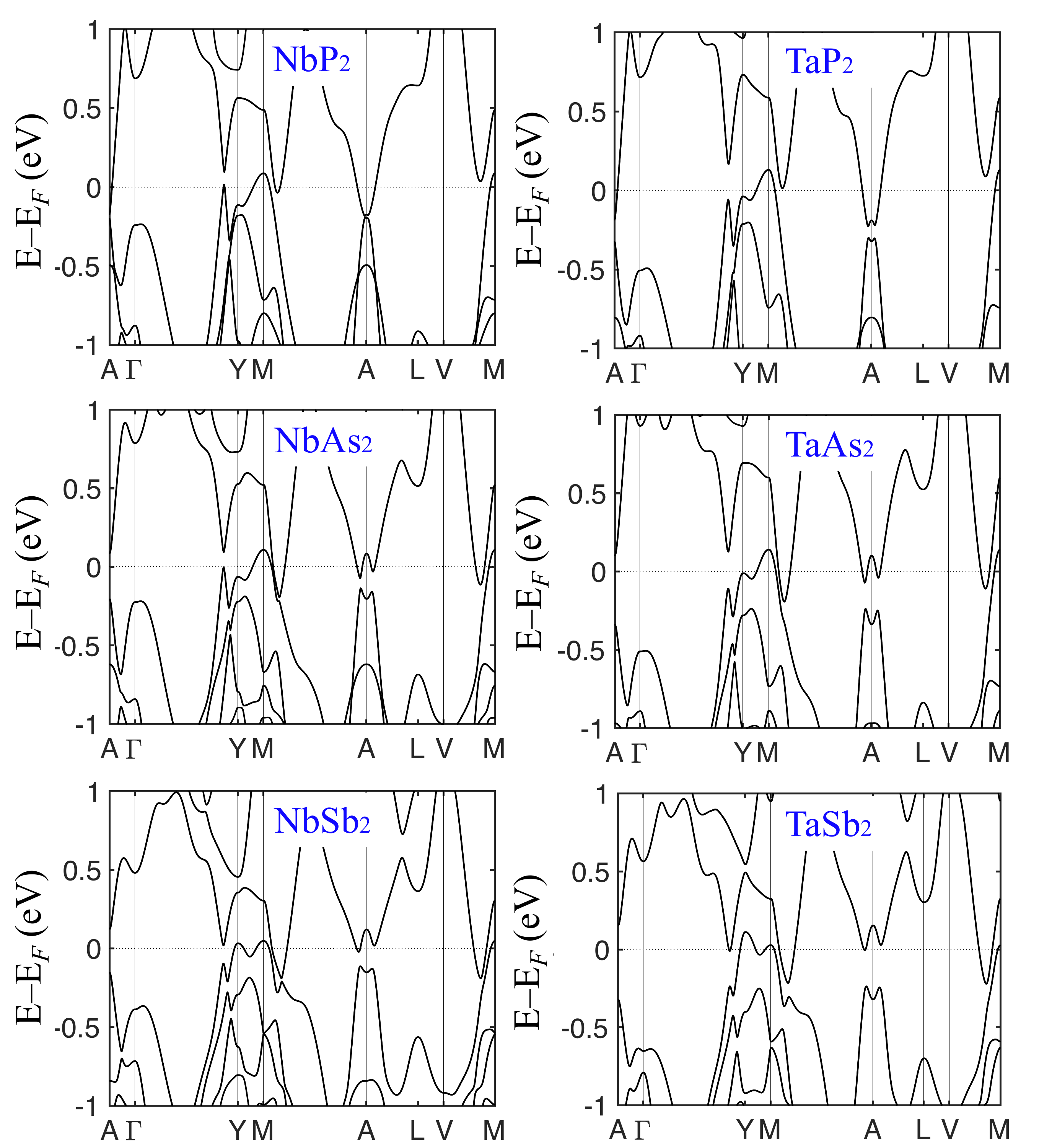}
\caption{Same as Fig. \ref{bs_nosoc} except that here the SOC is included. The gapped nodal lines can be seen along the $\Gamma-Y$ and $M-A$ directions.}
\label{bs_soc}
\end{figure}

In Figs. \ref{bs_nosoc} and \ref{bs_soc}, we present band structure of various members of the transition metal dipnictide ${\rm RX_2}$ family without and with spin-orbit coupling (SOC), respectively.  We consider six members of this family, which are listed in Table \ref{tab:t1} along with the lattice parameters used in the calculations. All these materials exhibit band crossings along the $\Gamma-Y$ and $M-A$ directions, which form nodal lines in the Brillouin zone (BZ) without the SOC similar to the case of $\rm TaAs_2$. Inclusion of SOC, gaps out the nodal lines and separates the valence and conduction bands locally at each $k$-point, and yields the $C_{2[010]}$-protected TCI states. The size of the inverted bandgap increases in the order P $<$ As $<$ Sb and Nb $<$Ta, which is consistent with the increasing order of the intrinsic SOC strength of the constituents atoms. We find that all materials have the same symmetry indicators with $(Z_2Z_2Z_2;Z_4) = (111; 2)$ and that the associated topological invariants are similar to those of $\rm TaAs_2$.

\begin{table}[!htb]
\centering
\caption{Experimental lattice constants $a$, $b$, $c$, and $\beta$ and relaxed-internal ionic positions of six members of the $RX_2$ materials family  \cite{Wang:2014nbsb2, Li:2016tasb2, Wang:2016nbas2,Wu:2016apl, Rundqvist:1966new}.}
\begin{ruledtabular}
\begin{tabular}{ccccccc}
 &${\rm NbP_2}$ & ${\rm NbAs_2}$ & ${\rm NbSb_2}$ & ${\rm TaP_2}$ & ${\rm TaAs_2}$ & ${\rm TaSb_2}$ \\
 \hline
 $a($\AA$)$ & 8.872 & 9.354 & 10.233 & 8.861 & 9.329 & 10.233 \\

  $b($\AA$)$ & 3.266 & 3.381 & 3.630 & 3.268 & 3.385 & 3.645 \\
  
  $c($\AA$)$ & 7.510 & 7.795 & 8.328 & 7.488 & 7.753 & 8.292 \\
  
  $\beta$ & 119.10 & 119.40 & 120.04 & 119.31 & 119.70 & 120.39 \\

  $x_R$ & 0.154 & 0.154 & 0.157 & 0.154 & 0.157 & 0.16  \\

  $z_R$ & 0.200 & 0.196 & 0.196 & 0.200 & 0.196 & 0.20  \\

  $x_{X1}$ & 0.399 & 0.399 & 0.404 & 0.399 & 0.406 & 0.40 \\

  $z_{X1}$ & 0.112 &  0.107 & 0.196 & 0.112 & 0.107 &  0.11 \\

  $x_{X2}$ & 0.143 & 0.140 & 0.142 & 0.143 & 0.139 & 0.14. \\

  $z_{X2}$ & 0.531 & 0.526 & 0.527 & 0.531 & 0.526 & 0.53
\end{tabular}
\end{ruledtabular}
\label{tab:t1}
\end{table}

\end{document}